\documentclass[12pt]{article}
\usepackage{docident}
\usepackage{latexsym}

\title{Redefintions of Histories by Measurements - An Explanation of
``Nonlocality'' Observed in EPR-Bohm Experiments}
\author{Yuri F. Orlov}
\begin{document}

\maketitle
\docident{\hspace{\fill}\makebox[0pt][r]{\sf CLNS 97/1475}}

\begin{center}
Floyd R. Newman Laboratory of Nuclear Studies\\
Cornell University, Ithaca, New York 14853 USA

%\medskip

{\bf Abstract}
\end{center}

It is proved in the frame of standard quantum mechanics that selection
of different ensembles emerging from measurements of an observable
leads to identification of corresponding reductions of the initial,
premeasured state.  This solves the problem of ``nonlocality``
observed in EPR-Bohm-type experiments.

%\medskip

The phenomenon of ``nonlocality`` is well-established in EPR-Bohm-type
experiments (EPRB).  For example, in entangled two-particle systems
[1,2], a choice of measurement in one of the two causally separated
channels influences results of measurements in another channel.  The
observed correlations are predicted by quantum mechanics (QM); but
since measurement procedures are not included in its formalism, their
mutual, long-distance influence remains inexplicable.  This difficulty
is cleared up in this work by analyzing logical statements about
quantum states in the frame of the standard QM formalism, and by
clarifying the meaning of measurement procedures.  This approach
permits us to find logical connections otherwise hidden.  It is
proved that in EPRB all such connections are {\it local.}  In general,
the choice of measurement in one of the channels locally defines the
reduction of a {\it premeasured} state, which in turn defines
connections between channels in the usual way.

%\medskip  

Let $\vert k_{l}\rangle, l = 1,2,\cdots,$ be eigenstates of an
observable(or of a number of commuting observables), K, considered
here as discrete, with all $k_{l}$'s different, and let
$\vert\Psi\rangle$ be an arbitrary state of a physical system,
$\vert\Psi\rangle={\sum}_{l}a_{l}\vert k_{l}\rangle$.  An exact
measurement of K with result $K=k_{i}$ leads to the reduction
$\vert\Psi\rangle\Rightarrow\vert k_{i}\rangle$.  It is generally
accepted, without proof, that reduced states describe only
developments of physical systems {\it subsequent} to measurements.
Below, in Theorem 3, we prove that under special conditions a
reduction of a state may act in both directions of time, so one can
replace premeasured state $\vert\Psi\rangle$ by its reduced part,
using the information received from the measurement.

%\medskip

{\it Lemma 1.}  The truth of a logical statement about a numerical
value of a physical observable is itself an observable, and is
represented in quantum mechanics by the pure state density matrix
or by a sum of such matrices.

%\medskip

{\it Proof.} Consider the logical statement $\Lambda_{k_{l}}$: ``The
system is in state $\vert k_{l}\rangle$`` or, in short, ``$K=k_{l}$.``
$\Lambda_{k_{l}}$ can be either ``true`` (the numerical value 1) or
``false`` (the value 0); these are two values of the classical logical
{\it truth of statement} $\Lambda_{k_{l}}$.  Statements corresponding
to different $k_{j}$'s (let us call them ``elementary``) are mutually
exclusive.  $\Lambda_{k_{l}}$ is true if the measurement of {\it K}
results in $k_{l}$, and false otherwise.  Therefore, $\Lambda_{k_{l}}$
is an observable and, like any other observable in QM, should be
represented by a Hermitian operator, $\hat{\Lambda}_{k_{l}}$, the
``truth operator of statement $\Lambda_{k_{l}}$,''$\left(
\hat{\Lambda}_{k_{l}}\right) ^{2}=\hat{\Lambda}_{k_{l}}$.  $\hat{K}$
and $\hat{\Lambda}_{k_{l}}$ have a common set of eigenvectors, $\vert
k_{j}\rangle, j=1,2,\cdots; \hat{\Lambda}_{k_{l}}\vert
k_{j}\rangle=\delta_{lj}\vert k_{j}\rangle$.  Here $\delta_{lj}$ is an
eigenvalue: $\Lambda_{k_{l}}$=1 (``true``) if $j=l$, and
$\Lambda_{k_{l}}$=0 (``false``) otherwise.  We conclude that
$\hat{\Lambda}_{k_{l}}$ must be identified with the density matrix of
the pure state $\vert k_{l}\rangle:\hat{\Lambda}_{k_{l}}\equiv\vert
k_{l}\rangle\langle k_{l}\vert$.  Since our choice of {\it K} was
arbitrary, a density matrix of any pure state $\vert\Psi\rangle$
defined in the Hilbert space of the physical system,
$\hat{\Lambda}_{\psi}\equiv\vert\Psi\rangle\langle\Psi\vert,
\hat{\Lambda}_{\psi}\vert\Psi\rangle =\vert\Psi\rangle,$ and
$\hat{\Lambda}_{\psi}\vert\overline{\Psi}\rangle$=0, where
$\vert\overline{\Psi}\rangle$ is any state orthogonal to
$\vert\Psi\rangle$, represents the logical statement, ``The system is
in state $\vert\Psi\rangle$.``  The truth of such a statement depends
on whether the physical system is really in state $\vert\Psi\rangle$.

%\medskip

It is easy to show [3,4] that every nonelementary statement about
numerical values of commuting physical observables is represented by a
corresponding sum of density matrices.  If there is a degeneracy such
that the same state corresponds to $k_{l_{1}},k_{l_{2}},\cdots$, then the
statement that the system is in this state is represented by
$\hat{\Lambda}_{k_{l_1}} + \hat{\Lambda}_{k_{l_2}}+\cdots\;\;\Box$ 

%\medskip

{\it Theorem 1. On the existence of exact locations.}  If the state of
the system, $\vert\Psi\rangle,$ and its observable, {\it K}, are such
that $\vert\Psi\rangle$ can be written as a superposition
\begin{equation}
\vert\Psi\rangle = \sum^{ln}_{l_{1}}c_{l}\vert k_{l}\rangle,
\end{equation}
then the system is located at some point $k_{i}$ of K-space, provided
that $\vert k_{i}\rangle$ is represented in the superposition.

%\medskip

{\it Proof.}  Let us find the operator of the following statement
about the numerical value of {\it K}:
\begin{equation}
\Lambda(\Psi,K): \Lambda_{k_{l_1}}\vee\Lambda_{k_{l_2}}\vee\cdots\vee\Lambda_{k_{l_n}}\equiv(``K=k_{l_{1}}``)\vee(``K={k_{l_2}}``)\vee\cdots\vee(``K={k_{l_n}}``),
\end{equation}
in which only those $k_{l}$'s that are in the superposition are
represented.  $\Lambda(\Psi,K)$ is symmetric relative to permutations
of its constituent mutually exclusive elementary statements.  Using
Lemma 1, the condition of mutual exclusiveness in the operator
representation,
$\hat{\Lambda}_{k_{l}}\hat{\Lambda}_{k_{m}}=\delta_{lm}\hat{\Lambda}_{k_{l}}$,
and equation
$\left[\hat{\Lambda}(\Psi,K)\right]^{2}=\hat{\Lambda}(\Psi,K)$, we find  
\begin{equation}
\hat{\Lambda}(\Psi,K)=\sum^{ln}_{l_{1}}\hat{\Lambda}_{k_{l}}\equiv\vert
k_{l_{1}}\rangle\langle k_{l_{1}}\vert +\vert k_{l_{2}}\rangle\langle
k_{l_{2}}\vert +...+\vert k_{l_{n}}\rangle\langle k_{l_{n}}\vert.
\end{equation}
From this
\begin{equation}
\hat{\Lambda}(\Psi,K)\vert\Psi\rangle =\vert\Psi\rangle,
\end{equation}
i.e., when the system is in state $\vert\Psi\rangle, \Lambda(\Psi,K)$
is a true statement.  Thus, according to the meaning of
$\Lambda(\Psi,K),$ this system is located in {\it K}-space at one of
the points enumerated in (2). 

%\medskip

{\it Corollary.}  The statement-disjunction,
$\Lambda_{k_{1}}\vee\Lambda_{k_{2}}\vee\cdots,$ containing {\it all}
possible numerical values of {\it K}, is true in {\it every}
state-superposition (1).

%\medskip

From Theorem 1 it follows, for example, that when a (nonrelativistic)
system is in a state of a certain momentum, $\vert p\rangle$, it is
also certain that this system is located {\it somewhere} in the
coordinate {\it q}-space.  Such a statement does not contradict QM,
but where this location is remains uncertain.  The next theorem shows
why such a question has no logical meaning.

%\medskip

{\it Theorem 2.} If {\it K} and {\it L} are two noncommuting
observables, and the state (1) of system, $n > 1,$ is an
eigenstate of operator $L, L\vert\Psi\rangle =
l\vert\Psi\rangle$, then there does not exist a logical statement,
either true or false, about the exact location of this system in {\it
K}-space.

%\medskip

We will omit the formal proof of this theorem.  Nor will we discuss
here the origins of noncommutativity.  If noncommutativity is granted,
then Theorem 2 provides the basis for quantum indeterminism
\cite{Orlov}: in the general case, we cannot describe with certainty
the future of the system undergoing a measurement.  The following
theorem states that this can be incorrect for the past.

%\medskip

{\it Theorem 3. On the redefinition of history by measurements.}  Let
{\it K} be an observable such that, in Heisenberg's representation,
the commutator, $\hat{K}(t-\delta
t)\hat{K}(t)-\hat{K}(t)\hat{K}(t-\delta t)\to 0$ when the time interval
$\delta t\to 0.$ (In particular, $\hat{K}$ may not depend on time at
all.)  Let the measuring procedure satisfy the condition of ``ideal
measurement`` formulated below.  And let a single measurement at time
$t_{0}$, by definition, result in some {\it K=k} only when statement
$(\Lambda_{k})_{t_{0}msr}$ about that measured value is formulated,
and the case separated from cases $k \neq k$; the corresponding
postmeasured ensemble will be called ``ensemble {\it K = k}.`` The
following logical implication is true: If a measured state is a
superposition, $\vert\Psi\rangle = \sum_{l}c_{l}\vert k_{l}\rangle$,
and a single measurement results in $K = k_{j}$ at time $t_{0}$, then
{\it after} this measurement, the statement: ``{\it Before} $t_{0}$,
the system belonging to the ensemble $K = k_{j}$ was in state $\vert
k_{j}\rangle$`` is true.

%\medskip

{\it Proof.}  First we formulate a condition of an ``ideal
measurement,`` a postulate that is commonly assumed and does not
contradict practice.  We assume that an observable, {\it K}, is being
measured.  If the measured system is in state $\vert k_{l}\rangle, l
=1$, or 2, or 3,$\cdots$, then the result of the measurement is $K=k_{l}$.
We will write this condition as the following logical implication
(valid for every {\it l}):
\begin{equation}
\left(\Lambda_{k_{l}}\right)_{t_{0}-\delta t}\to \left(\Lambda_{k_{l}}
\right)_{t_{0}msr}
\end{equation}
Statement (5) makes sense since its constituent statements, being
formalized in QM, are assumed to commute, at least in the
approximation $\delta t\to 0$; otherwise, according to Theorem 2, it
would be meaningless.  From (5) it follows that if the premeasured
system is in one of the states $\vert k_{j}\rangle$ such that $j\neq
l$, i.e., in any eigenstate of {\it K} but $\vert k_{l}\rangle$, then
the result of the measurement of {\it K} is a $k_{j},j\neq l$.  This
can be written as
\begin{equation}
\left(\Lambda_{k_{1}}\vee\Lambda_{k_{2}}\vee\cdots\vee\Lambda_{k_{l-1}}\vee
\Lambda_{k_{l+1}}\vee\cdots\right)_{t_{0}-\delta t}\to
\left(\Lambda_{k_{1}}\vee\cdots\vee\Lambda_{k_{l-1}}\vee\Lambda_{k_{l+1}}\vee
\cdots\right)_{t_{0}msr}
\end{equation}
The premise is true if at least one of the elementary statements on
the left side is true, and this is indeed the case; then according to
(5), that very same elementary statement on the right side is true
also, so the conclusion is also true.  Now, according to the corollary
of Theorem 1, the disjunction of {\it all} possible elementary
statements $\Lambda_{k_{l}}, l = 1,2,\cdots,$ about numerical values of
{\it K} is always a true statement (a tautology).  Therefore, the
disjunction from which only one elementary statement is excluded, as
it is in (6), is a logical equivalent (denoted by $\sim$) of the logical
negation of this statement:
\begin{equation}
\Lambda_{k_{1}}\vee\Lambda_{k_{2}}\vee\cdots\Lambda_{k_{l-1}}\vee
\Lambda_{k_{l+1}}\vee\cdots\sim\bar{\Lambda}_{k_{l}}
\end{equation}
Indeed, if one of the statements on the left side of the equivalence
is true, then the left side is true and all other statements about
numerical values of $K,\Lambda_{k_{l}}$ included, are false; therefore,
$\bar{\Lambda}_{k_{l}}$ on the right side is true.  If none of the
statements on the left side is true, then the left side is false and
$\Lambda_{k_{l}}$ is true, as only possibility left to the physical
system.  Therefore, $\bar{\Lambda}_{k_{l}}$ is false, as is the left
side.  As a result, we can rewrite (6) as
\begin{equation}
\left(\bar{\Lambda}_{k_{l}}\right)_{t_{0}-\delta t}\to
\left(\bar{\Lambda}_{k_{l}}\right)_{t_{0}msr}
\end{equation}
And finally, as logically follows from (8),
\begin{equation}
\left(\Lambda_{k_{l}}\right)_{t_{0}msr}\to
\left(\Lambda_{k_{l}}\right)_{t_{0}-\delta t}
\end{equation}

%\medskip

Implication (9) can be used to conclude whether the state of the
system was really $\vert k_{l}\rangle$, only when it is certain that
$\left(\Lambda_{k_{l}}\right)_{t_{0}msr}$ is true, that is, when the
results of the measurement formulated as ``$K=k_{l}$`` have been
selected.  Absent this procedure, the detector is not being used as a
measuring device but only as a target.  In such a case, some quantum
state emerges as a result of the interaction of the system with this
target, so premise $\left(\Lambda_{k_{l}}\right)_{t_{0}msr}$ in (9) is
not true.

%\medskip

For the ensemble selected as $K=k_{l}$, the a posteriori conclusion
from (9) and from the results of measurement is that the state of the
system before the measurement was $\vert k_{l}\rangle$[5,6]. $\Box$ 

\medskip

{\it Corollary.}  If an observable {\it K} satisfies the commutation
conditions of Theorem 3, then when we select an ensemble corresponding
to a result {\it K=k} of our measurement, we simultaneously select an
ensemble corresponding to the value {\it K=k} of {\it premeasured}
physical systems.  (Note that the collapse of the premeasured state
accompanying this selection is a purely informational effect.)

%\medskip

In EPRB [1,2,7-9] the observables are polarizations of particles,
their operators do not depend on time, and $\delta t$ may be finite.
In such an experiment, let a pair of particles be prepared in an
entangled state $\vert\Psi\rangle$ at time {\it t}=0 (for the center
of their wave packet), and the time-size of the wave packet,
$\triangle t\sim\hbar/\triangle E$, be much less than the flight time
to either of two detectors.  Let detector D1 measure observable {\it K} of
particle 1 moving in channel 1, and {\it L} be an observable of
particle 2 in channel 2 with its detector, D2.  In the general case,
\begin{equation}
\vert\Psi\rangle =\sum_{ij}a_{ij}\vert k_{i}\rangle_{1}\vert l_{j}\rangle_{2},
\end{equation}
where 1(2) refers to particle 1(2), and $k_{i},l_{j}$ are eigenvalues
of {\it K,L.} The phenomenon called ``nonlocality`` is the influence
of random choices of noncommuting observables $K, K', K''$, $\cdots$, 
to measure, say, particle 1 in channel 1, on the
outcomes of independent measurements of particle 2 in the other
channel.  The significance of Theorem 3 is that it permits establishing
deterministic logical connections, as in classical physics, between
postmeasured states of particle 1 selected in channel 1, and
premeasured states of particle 2 that are thereby selected also.  Let
observable {\it K} be measured at time $t_{0}$, and the postmeasured
ensemble $K=k_{n}$ selected. Then it can be concluded that the
premeasured state of particle 1 in ensemble $K=k_{n}$ is $\vert
k_{n}\rangle_{1}$.  But in common state (10), $\vert k_{n}\rangle_{1}$
is coupled one-to-one with $\vert
l'_{m}\rangle_{2}=A\sum_{j}a_{nj}\vert l_{j}\rangle_{2}$, where
$1/\vert A\vert^{2}=\left(w_{k_{n}}\right)_{1}$ is the probability of
a $(K=k_{n})$ result in channel 1, and $l'_{m}$ a value of an
observable $L'$.  Therefore, the state of coupled particle 2 is
$\vert l'_{m}\rangle_{2}$.  By selecting the one-particle, {\it
post}measured ensemble $K=k_{n}$, then, we have also automatically
selected the two-particle, {\it pre}measured ensemble described by the
reduced state $\vert k_{n}\rangle_{1}\vert l'_{m}\rangle_{2}.$  Thus,
there are no nonlocal physical influences; ``nonlocality`` in the
sense of one-to-one correspondence between selected states of the
spatially separated particles, $k_{n}\leftrightarrow l'_{m}$, is a
result of the common history of the two particles.

%\medskip

It is interesting that in EPRB we can even reconstruct the logical
chain, with every infinitesimal link local, from the measurement in
channel 1 to the state of particle 2 in channel 2.  Since observables
{\it K} and {\it L} in EPRB are constant before measurements, we
conclude that at time $t_{0}=N\delta t$ the value of {\it K} is the
same as at time $t_{0}-(N-1)\delta t, N=1,2,\cdots$. From Theorem 3 we
know that at time $t_{0}-\delta t, K=k_{n}$.  Therefore, at time $t=0$,
when the pair of particles is born, the state of particle 1 in
ensemble $K=k_{n}$ is $\vert k_{n}\rangle_{1}$.  Therefore, the state
of particle 2 at that moment is $\vert l'_{m}\rangle_{2}$.  Applying
similar logical steps to channel 2, we conclude that particle 2
conserves its state $L'=l'_{m}$ until some measurement in channel 2 is
made either before or after $t_{0}$.

%\medskip

Let detector D2 now measure an observable, {\it L}.  If {\it L}
commutes with $L'$, then according to (5) the result of this
measurement is deterministic and equals $l=l'_{m}$.  If {\it L} does
not commute with $L'$, the transition
$\vert l'_{m}\rangle_{2}\rightarrow\vert l_{j}\rangle_{2}$, where
$l_{j}$ results from the measurement of {\it L}, is unpredictable.
Applying Theorem 3, now, not to the measurement of {\it K} but to the
measurement of {\it L} by detector D2, and selecting postmeasured
ensemble $L=l_{j}$, we will thereby automatically select a
two-particle, premeasured ensemble different from the two-particle
ensemble connected with ensemble $K=k_{n}$.  Indeed, the result of
measurement in channel 2, $L=l_{j}$, defines the state of particle 1
in channel 1, $\vert k'_{p}\rangle_{1}=B\sum_{i}a_{ij}\vert
k_{i}\rangle_{1}$, where $k'_{p}$ is a value of an observable $K'$.
From this we can conclude that the premeasured reduced state of
the two-particle ensemble is $\vert k'_{p}\rangle_{1}\vert
l_{j}\rangle_{2}$ . Despite the differences between ensembles,
however, probability
$w\left(k_{n},l_{j}\right)=\langle\Psi\vert\hat{\Lambda}_{k_{n}}\hat{\Lambda}_{l_{j}}\vert\Psi\rangle=\vert
a_{nj}\vert^{2}$ for measurements in both channels does not depend on
which of the two possible ensembles, $\vert k_{n}\rangle_{1}\vert
l'_{m}\rangle_{2}$ or $\vert k'_{p}\rangle_{1}\vert l_{j}\rangle_{2}$,
is chosen.  Our a posteriori conclusion about the collapse of initial state
$\vert\Psi\rangle$ therefore has an uncertainty caused by
noncommutativity, and depends on the information we choose to be our
premise --- either $``K=k_{n}``$ or $``L=l_{j}``$ (or both).


\begin{thebibliography}{9}
\bibitem{Mistry} Nariman B. Mistry has proposed a hypothesis similar in some
                   respects to Theorem 3 (personal communication).
\end{thebibliography}
\end{document}